\documentclass[conference]{IEEEtran}
\IEEEoverridecommandlockouts
\usepackage{amsmath,amssymb,amsfonts}
\usepackage{algorithmic}
\usepackage{graphicx}
\usepackage{textcomp}
\usepackage{xcolor}
\def\BibTeX{{\rm B\kern-.05em{\sc i\kern-.025em b}\kern-.08em
    T\kern-.1667em\lower.7ex\hbox{E}\kern-.125emX}}
\usepackage[utf8]{inputenc}
\usepackage{datetime}
\usepackage{upgreek}
\usepackage{dblfloatfix}    
\usepackage{subfigure}

\usepackage{multirow}
\usepackage{booktabs}
\usepackage[font=small]{caption}
\usepackage{xcolor,soul,lipsum}
\usepackage{bm}
\usepackage{makecell} 
\usepackage{xr-hyper}
\usepackage{hyperref}
\hypersetup{
    colorlinks=true,
    linkcolor=blue,
    filecolor=magenta,      
    urlcolor=cyan,
}

\usepackage{threeparttable}
\makeatother
\usepackage[hyphenbreaks]{breakurl}
\usepackage[acronym]{glossaries}
\makeatletter
\usepackage[style=ieee, backend=bibtex, defernumbers=true]{biblatex}
\DeclareSourcemap{
  \maps[datatype=bibtex]{
    \map[overwrite]{
      \step[fieldsource=doi, final]
      \step[fieldset=url, null]
      \step[fieldset=eprint, null]
    }  
  }
}
\usepackage{upgreek} 

\usepackage[capitalize]{cleveref}
\crefname{section}{Sec.}{Secs.}
\Crefname{section}{Sec.}{Sects.}
\Crefname{table}{Table}{Tables}
\crefname{table}{Table}{Tabs.}

\begin{document}

\title{Exploiting Alternating DVS Shot Noise Event Pair Statistics to Reduce Background Activity Rates\\ \thanks{RG supported by Swiss National Science Foundation grant SCIDVS (200021\_185069).} }

\author{\IEEEauthorblockN{Brian McReynolds, Rui Graca, Tobi Delbruck}
\IEEEauthorblockA{\textit{Sensors Group, Inst. of Neuroinformatics, UZH-ETH Zurich, 
Zurich, Switzerland} \\
bmac,rpgraca,tobi@ini.uzh.ch, \url{https://sensors.ini.uzh.ch}\\
}
}

\maketitle

\begin{abstract}
Dynamic Vision Sensors (DVS) record "events" corresponding to pixel-level brightness changes, resulting in data-efficient representation of a dynamic visual scene. As DVS expand into increasingly diverse applications, non-ideal behaviors in their output under extreme sensing conditions are important to consider. Under low illumination (below $\approx$10 lux) their output begins to be dominated by shot noise events (SNEs) which increase the data output and obscure true signal. SNE rates can be controlled to some degree by tuning circuit parameters to reduce sensitivity or temporal response bandwidth at the cost of signal loss. Alternatively, an improved understanding of SNE statistics can be leveraged to develop novel techniques for minimizing uninformative sensor output. We first explain a fundamental observation about sequential pairing of opposite polarity SNEs based on pixel circuit logic and validate our theory using DVS recordings and simulations. Finally, we derive a practical result from this new understanding and demonstrate two novel biasing techniques shown to reduce SNEs by 50$\%$ and 80$\%$ respectively while still retaining sensitivity and/or temporal resolution.          
\end{abstract}

\label{sec:abstract}

\newacronym{DVS}{DVS}{Dynamic Vision Sensors}
\newacronym{ISI}{ISI}{Inter-Spike Interval}
\newacronym{SNR}{SNR}{Signal to Noise Ratio}
\newacronym{SNE}{SNE}{Shot Noise Event}

\newcommand{\vpd}{V_\text{pd}}

\section{Introduction}

\gls{DVS}, or event cameras, efficiently encode dynamic visual information into a sparse stream of ON (increasing brightness) and OFF (decreasing) events with high temporal resolution. This sensing paradigm has several benefits including wide dynamic range, high temporal resolution, and low power consumption. 
\gls{DVS} have already proven useful for many applications related to machine vision \cite{Gallego2020-pa}. Despite these benefits, physical noise sources cause erroneous events even when there are no brightness changes in the scene, and elevated noise rates when illumination is low have thus far hindered widespread adoption in applications requiring high performance in dim lighting. 

Under low illumination, \gls{SNE}s dominate \gls{DVS} noise~\cite{Hu2021-v2e-ieee, Suh2020-samsung-dvs, Finateu2020-prophesee-isscc}, and denoising \gls{DVS} output has been the focus of numerous efforts \cite{Khodamoradi2017-ke,Guo2022-am,Czech2016-by}. Although many custom denoising strategies have been developed, none explicitly consider noise event-pair statistics. Many aspects of \gls{DVS} noise remain difficult to predict, but recent work has made significant progress toward understanding of the processes and trade-offs that influence \gls{SNE}s \cite{graca2021unravelingtheparadox,delbruck2021feedbackcontrol,graca2023shininglight}. 

We expand on these efforts and explain a simple yet previously unreported behavior inherent to the self-timed reset necessary for \gls{DVS} pixel operation. In ~\cref{sec:dvs_operation} we describe the basic functionality of the \gls{DVS} pixel with an emphasis on the circuit behavior that influences noise statistics.  ~\cref{sec:event_pairs} describes the observation that \gls{SNE}s tend to occur in opposite polarity (ON/OFF) pairs, and explains this behavior based on pixel reset logic. ~\cref{sec:bias_adjustments} then demonstrates a practical result of this observation by demonstrating two sensor bias techniques that reduce \gls{SNE} rates by directly manipulating noise statistics.              

\section{DVS Pixel Operation}
\label{sec:dvs_operation}
\begin{figure*}[h]
    \centering
    \includegraphics[width=\textwidth]{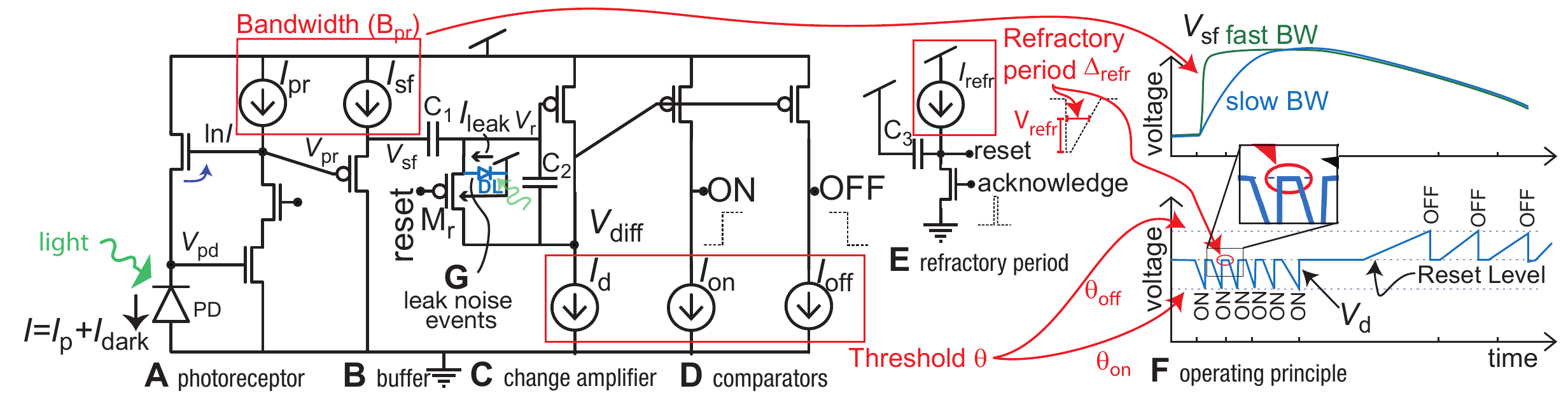}
    \caption{Typical DVS pixel circuit schematic. The active logarithmic photoreceptor front-end (\textbf{A-B}) drives a cap-feedback change amplifier (\textbf{C}) with output \textbf{$V_{diff}$}. When \textbf{$V_{diff}$} deviates by either an ON or OFF threshold, comparators (\textbf{D}) report an event, and after a finite refractory period (\textbf{E}) the change amplifier is reset. \textbf{F:} After each reset, the pixel again responds to signal changes around a new reference level. This reset logic explains the key observation of this paper.}
    \label{pixel-circuit}
\end{figure*}

The first practical \gls{DVS} pixel was introduced in \cite{Lichtsteiner2008-dvs}, and modern event camera pixels are based on the same fundamental stages described in ~\cref{pixel-circuit}.  These core components are a logarithmic transimpedance photoreceptor which generates an output voltage, $V_{pr}$, proportional to log photocurrent, a change amplifier that amplifies signal changes around a fixed reference point, two independent comparators for generating ON and OFF output events when the signal changes by a tunable threshold value, and a circuit to reset the change amplifier after each event to allow the pixel to respond to changes around a new reference level. In most cases, this new reference is approximately the signal level that generated the previous event. Readout circuits in the focal plane periphery record and timestamp the resulting sequence of ON and OFF events to encode pixel-level brightness changes.  

Pixel behavior is refined by adjusting programmable biases (highlighted in red and depicted as current sources in ~\cref{pixel-circuit}), allowing the user to tune performance for a variety of sensing tasks. $I_{pr}$ and $I_{sf}$ adjust the temporal response of the photoreceptor, which is also limited by the background photocurrent. The effects of these two biases on \gls{SNE} rates are extremely complex, and thoroughly described in \cite{graca2023optimalbiasing}. The next set of biases define the independent ON and OFF thresholds, $\theta_{ON}$ and $\theta_{OFF}$, which are proportional to $\log(\frac{I_{on}}{I_d})$ and $\log(\frac{I_{d}}{I_{off}})$ respectively. After each event, $M_r$ shorts the input and output of the change amplifier to prevent subsequent events during a refractory period or "dead-time", and opens again as the reset node rises. $I_{refr}$ controls the rate at which the reset node charges, and can be tuned to increase or decrease the maximum firing rate for individual pixels. Composite effects of these biases are further detailed in \cite{graca2023shininglight}. 

\section{Shot Noise Event Pairs}
\label{sec:event_pairs}

\begin{figure}[h]
    \centering
    \includegraphics[width=0.45\textwidth]{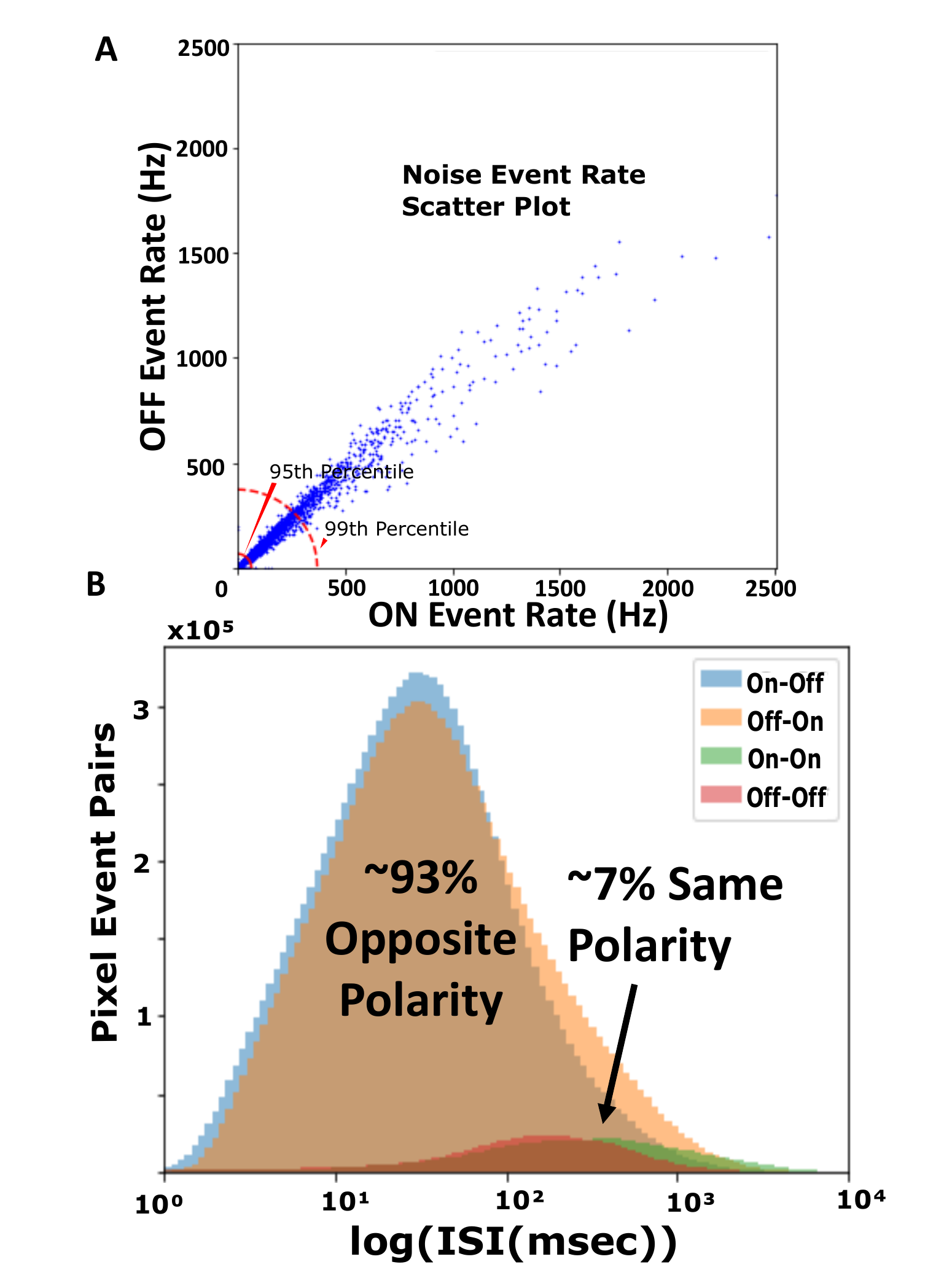}
    \caption{Recorded DAVIS346 \gls{SNE}s under 10 mlux illumination with high bandwidth biases. \textbf{A:} Per pixel ON and OFF \gls{SNE} rates are nearly balanced, even for pixels with an abnormally high noise rates. \textbf{B:} \gls{ISI} histograms reveal that over 90\% of pixel \gls{SNE} pairs are opposite polarity and occur at shorter time intervals than like polarity pairs.}
    \label{fig1}
\end{figure}

To better understand the root causes of \gls{DVS} \gls{SNE}s, examining the scatter plot of ON and OFF noise events shown in ~\cref{fig1} reveal ON and OFF events are nearly balanced in each pixel. At first glance, this result is counter-intuitive given the well-known mismatch in independent ON and OFF threshold levels \cite{Lichtsteiner2008-dvs,Finateu2020-prophesee-isscc}. Specifically, noise rates are known to increase dramatically with sensitivity \cite{graca2021unravelingtheparadox}. Because $\theta_{ON}$ and $\theta_{OFF}$ are independent, it is extremely unlikely that a pixel with a low $\theta_{ON}$ will also have an extremely low $\theta_{OFF}$. In ~\cref{fig1}, the 99th percentile is depicted by the outer dashed red arc, and ON and OFF \gls{SNE} rates of each type are still roughly balanced for pixels outside this region, indicating a dependency that is not explained by prior reasoning.  

To further explore and illustrate this phenomenon, we calculated the \gls{ISI} between consecutive event-pairs in each pixel and specifically examined the polarities of the pairs. Examining the \gls{ISI} distribution in ~\cref{fig1}B reveals that over 90\% of sequential noise event pairs are of opposite polarity \textbf{and} these pairs typically occur at shorter time intervals ($\approx 1/10$). Both of these observations about \gls{SNE} pairs are in contrast with previous assumptions, which predict noise events should be independent of pixel history. 

\begin{figure*}[h]
    \centering
    \includegraphics[height=6cm]{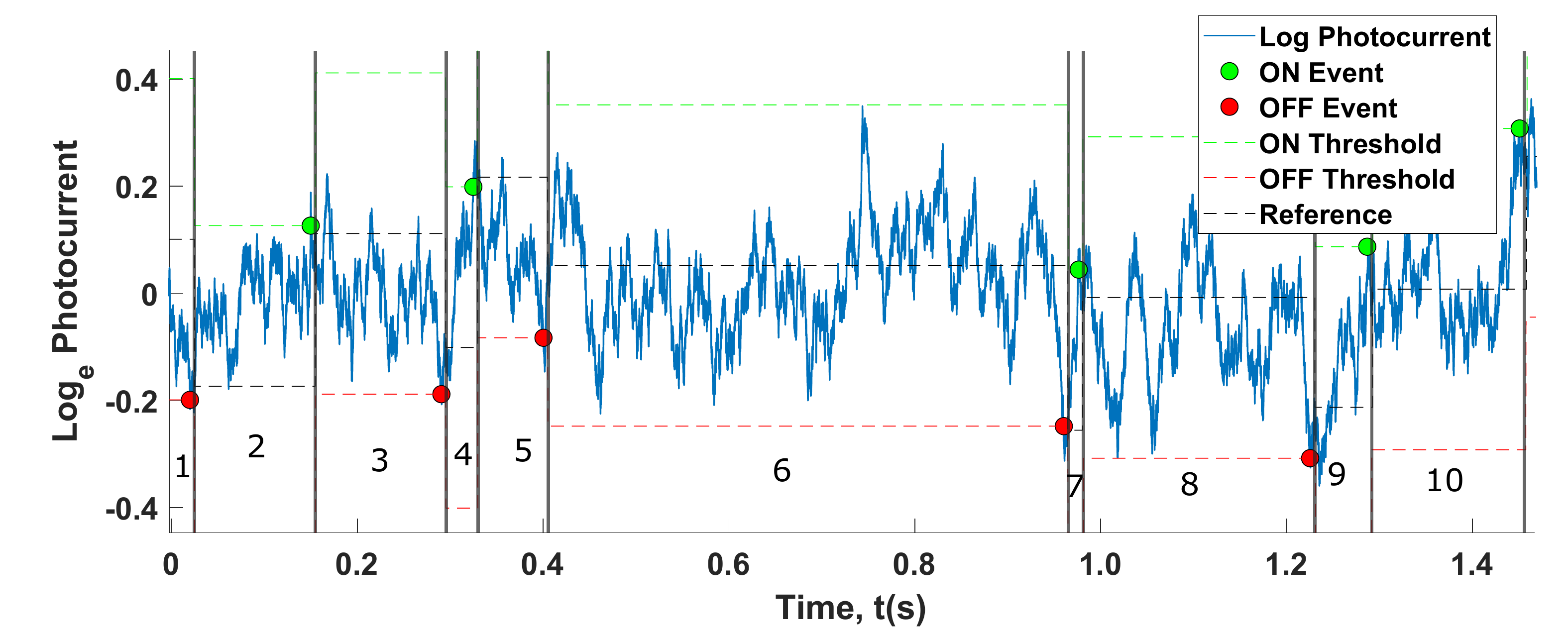}
    \caption{Explanation of noise event pairing.  Each time window (labeled 1-10) terminates with an event when the noisy signal crosses either the \textbf{ON} or \textbf{OFF} threshold.  Shortly after each event (dependent on refractory period), the reference level resets near the signal level that generated the previous event, increasing the probability that an opposite polarity noise event occurs. In the example, 8 of 10 event pairs are opposite polarity and occur on shorter time scales than like-polarity pairs.}
    \label{fig2}
\end{figure*}

\begin{figure}[h]
    \centering
    \includegraphics[width=0.45\textwidth]{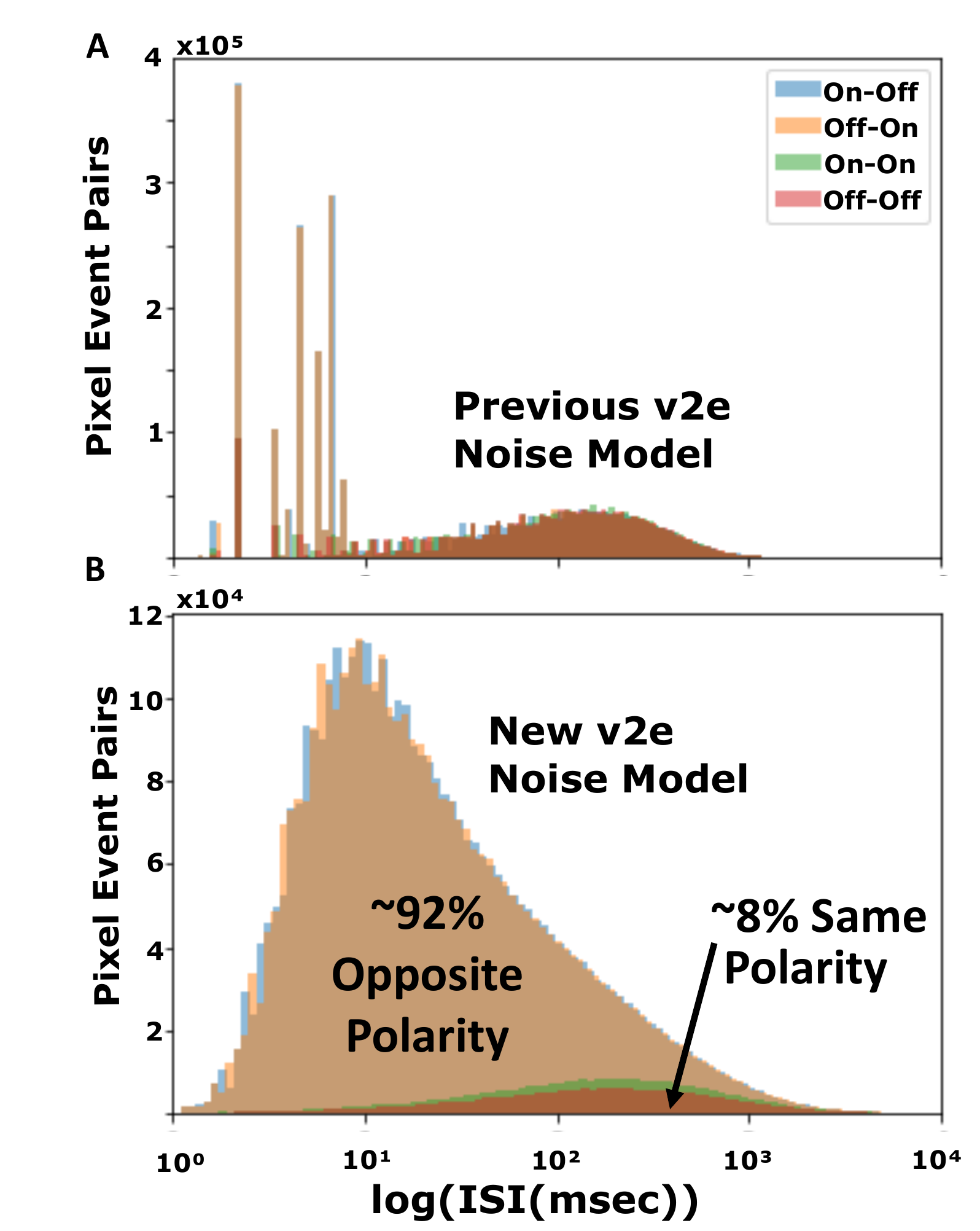}
    \caption{Comparison of old and new v2e~\cite{Hu2021-v2e-ieee} noise models.  \textbf{A:} The previous model did not accurately capture noise statistics.  \textbf{B:}  Adding Gaussian white noise to a DC signal and allowing the event generation model to generate noise events produces realistic \gls{DVS} noise statistics.}
    \label{fig3}
\end{figure}

~\cref{fig2} explains how this behavior is a direct result of the pixel’s self-timed reset. Events are generated when the signal deviates from a memorized reference level by more than an ON ($\theta_{ON}$) or OFF ($\theta_{OFF}$) threshold. Considering a filtered white gaussian noise pattern, each event resets the pixel’s reference to a level offset from the mean noise value. Since gaussian noise tends to return to its mean value, this new reference increases the probability of an event of opposite polarity happening within relatively short time. This hypothesis is upheld in \ref{fig3}, which demonstrates how improving the v2e \gls{DVS} simulator~\cite{Hu2021-v2e-ieee}\footnote{\href{https://github.com/SensorsINI/v2e}{v2e on github} - see \textsl{--photoreceptor\_noise} option} by injecting white noise prior to the event generation block accurately models observed noise statistics.

\section{Bias Adjustments for SNE Rate Reduction}
\label{sec:bias_adjustments}

When operating in dim conditions, noise rates are typically managed by reducing sensitivity or photoreceptor bandwidth, but true signal is suppressed as changes too small or fast for the selected biases are missed completely. If an application requires detecting fast moving or dim objects/features, moderately elevated noise rates can be accepted and aggressive denoising applied after reading events off-chip at the cost of increased latency, power, computation, and data bandwidth. Alternatively, reasoning from Fig.~\ref{fig2} reveals two novel biasing strategies to reduce background noise rates while still allowing pixels to be biased for high sensitivity and bandwidth.   

\begin{figure}[h]
    \centering
    \includegraphics[width=0.45\textwidth]{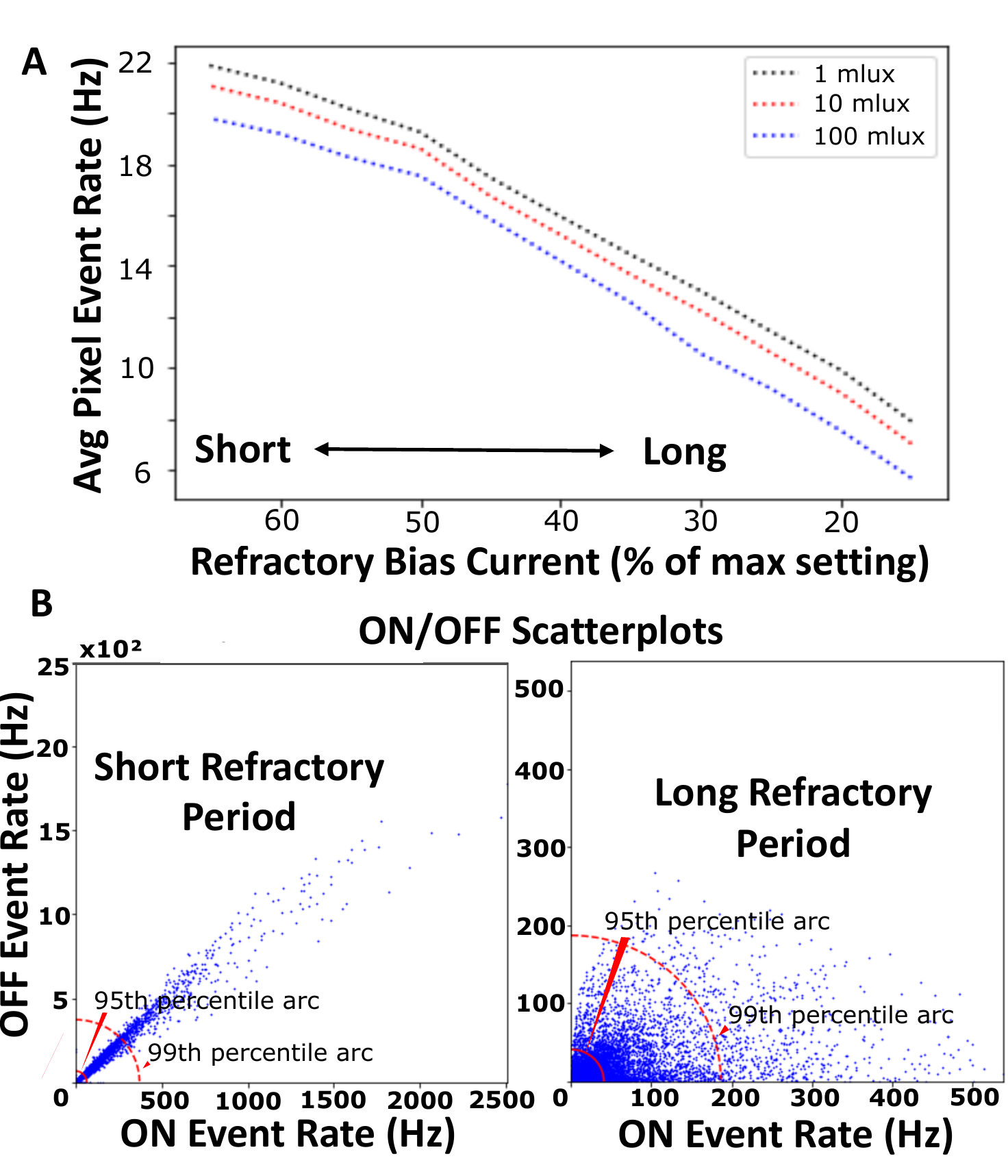}
    \caption{\textbf{A:} Noise measurements taken from a FSI DAVIS346 \cite{8334288} at three different illumination levels (1, 10 and 100 mlux). Horizontal axis is refractory period bias current ranging from 118pA (far left) to 8.7nA (far right). Increasing the refractory period reduces average noise rates by more than 50\% for all illumination levels.  \textbf{B:} Scatter plots of individual pixel ON and OFF noise events validate that the longer refractory period decouples ON and OFF event pairs.}
    \label{fig4}
\end{figure}

The first strategy is to increase the refractory period. ~\cref{fig4} demonstrates that this method decouples the reset level from the signal level that generated the previous noise event and reduces overall noise rates. ~\cref{fig4}A shows more than 50\% reduction in noise rates and ~\cref{fig4}B demonstrates decoupling of ON/OFF pairs with a longer refractory period. Simulations suggest that in order for this decoupling to occur, the refractory period must be $\geq\frac{1}{2\pi f_{3dB}}$, where $f_{3dB}$ is the low-pass corner frequency of the photoreceptor/source follower combination.         

\begin{figure}[h]
    \centering
    \includegraphics[width=0.45\textwidth]{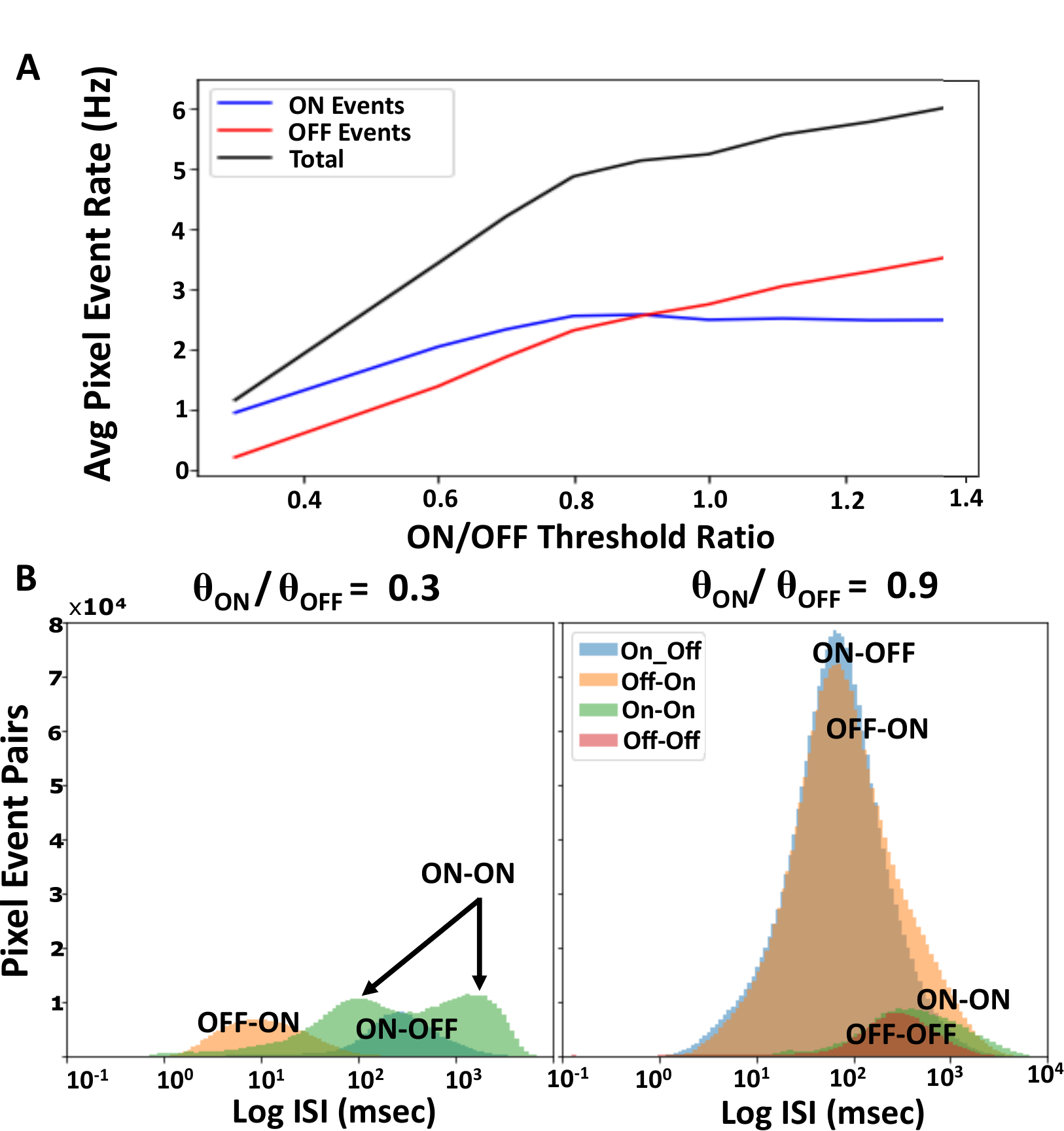}
    \caption{\textbf{A:} Noise measurements from a FSI DAVIS346 \cite{8334288} at 1 mlux for varied ratios of ON/OFF thresholds.  Total \gls{SNE} rates drop when the ON threshold ($\theta_{ON}$) is lower than the OFF threshold ($\theta_{OFF}$).  \textbf{B:} \gls{ISI} histograms show a shift in event-pair distributions, resulting in greater than 80\% reduction in overall noise rates when $\theta_{ON}/\theta_{OFF} \approx 0.30$.}
    \label{fig5}
\end{figure}

The second technique is deliberately applying a large imbalance in ON and OFF thresholds to force the reference level to settle near the extreme of the noise distribution corresponding to the more sensitive threshold. This large imbalance reduces the probability that a subsequent noise event will occur to reset the reference, thus breaking the event-pair cycle. In practice, ~\cref{fig5} shows that this method works well when ON is much more sensitive than OFF, and ~\cref{fig5}B demonstrates up to an 80\% reduction in noise event rates, even with an expected increase in sensitivity to ON changes.      

\section{Conclusion}

\gls{SNE} rate is an important consideration for expanding the utility of \gls{DVS} into diverse applications in challenging lighting conditions. In this paper, we identified a key observation about how \gls{SNE}s tend to occur in pairs of opposite polarity, and explained this phenomenon based on pixel architecture and logic. Leaning on this explanation, we propose and demonstrate two novel bias techniques for reducing \gls{SNE} rates. Limiting noise rates in dim lighting conditions improves \gls{DVS} \gls{SNR}, and the techniques we describe facilitate direct manipulation of noise statistics. Further exploration of the benefits of these techniques should be explored in task specific scenarios. After achieving desired \gls{SNR} performance, a deeper understanding of the resulting noise statistics can aid in more efficient and effective denoising strategies, and inform improvements to already effective machine learning-based denoisers such as~\cite{Guo2022-am}.

%

\renewcommand*{\bibfont}{\footnotesize}
\footnotesize{\printbibliography}



\end{document}